\documentclass[fleqn,10pt]{wlscirep}
\title{Thermal Sublimation: a Scalable and Controllable Thinning Method for the Fabrication of Few-Layer Black Phosphorus}

\author[1]{Weijun Luo}
\author[2]{Rui Yang}
\author[3]{Jialun Liu}
\author[1]{Yunlong Zhao}
\author[3]{Wenjuan Zhu}
\author[1,*]{Guangrui (Maggie) Xia}
\affil[1]{the University of British Columbia, Department of Materials Engineering, Vancouver, B.C., V6T1Z4, Canada}
\affil[2]{the University of British Columbia, Department of Physics and Astronomy, Vancouver, B.C., V6T1Z4, Canada}
\affil[3]{the University of Illinois at Urbana-Champaign, Department of Electrical and Computer Engineering, Champaign, IL, 61801, USA}

\affil[*]{gxia@mail.ubc.ca}
\usepackage{amsmath}
\usepackage{amssymb}
\usepackage{latexsym}
\usepackage{epstopdf}
\usepackage{graphicx}
\usepackage{dcolumn}
\usepackage{bm}
\usepackage{longtable}

\begin{abstract}
	In this work, we reported uniform layer-by-layer sublimation of black phosphorus under heating below 600 K. The uniformity and crystallinity of BP samples after thermal thinning were confirmed by Raman spectra and 2D Raman imaging. A uniform and crystalline bilayer black phosphorus flake with an area of 180 $\mu$m$^2$ was prepared with this method. No micron scale defects were observed. The sublimation rate of BP was around 0.18 nm / min at 500 K and 1.5 nm / min at 550 K. Both room and high temperature Raman peak intensity ratio $\frac{Si}{A_g^2}$ as functions of BP thickness were established for in-situ thickness determination. The sublimation thinning method was shown to be a controllable and scalable approach to prepare few-layer black phosphorus.
\end{abstract}
\begin{document}
	\flushbottom
	\maketitle
	
	\section{Introduction}
	Semiconducting orthorhombic black phosphorus (BP), like graphene, is a layer-stacked 2-dimensional (2D) nanomaterial with sp3-hybridized phosphorus atoms covalently bonded to a puckered structure via weak van der Waals forces\cite{ling2015renaissance} . 
	The energy band gap close to 2 eV has been measured by scanning tunneling spectroscopy (STS) from monolayer BP\cite{liang2014electronic} .
	With the increasing number of layers, the band gap decreases and finally reaches 0.3 eV for bulk BP\cite{takao1981electronic} . 
	This tunable band gap enables BP to be a promising candidate for applications in field effect transistors\cite{li2014black}$^,$\cite{kim2015dual}$^,$\cite{koenig2014electric}$^,$\cite{das2014ambipolar}$^,$\cite{du2014device} . 
	Other efforts have shown its prospective applications in gas sensors\cite{abbas2015black} , lithium ion batteries\cite{sun2012electrochemical} , thermoelectric devices\cite{qin2014hinge}$^,$\cite{lv2014large} , photo transistors\cite{low2014origin}$^,$\cite{wang2015ultrafast} .\\
	
	\begin{flushleft}
		So far, there remain 2 major challenges for BP research: air-stabilization and controllable fabricating ultra-thin BP layers \cite{castellanos2015black}. To address that, the passivation of BP was both successfully achieved via chemical modification method like covalent aryl diazonium functionalization \cite{ryder2016covalent} and physical encapsulation methods such as atomic layer deposited dielectric passivation or h-BN passivation \cite{luo2016surface}. However, current top-down methods towards ultra-thin BP such as mechanical exfoliation\cite{castellanos2014isolation}, anodic oxidation and water rinsing\cite{liu2016scanning}, shear exfoliation\cite{brent2014production}$^,$ \cite{hanlon2015liquid}$^,$\cite{seo2016triangular} in liquids, and plasma thinning\cite{lu2014plasma}$^,$\cite{lee2016tuning}$^,$\cite{pei2016producing} still couldn't be used for scalable and controllable production of large uniform few-layer phosphorene. In other aspect, uniformity, scalability and crystallinity of BP thin film made by bottom-up methods such as pulsed laser deposition\cite{smith2016growth}$^,$\cite{yang2015field} and chemical vapor deposition\cite{smith2016growth} (CVD) remained critical obstacles for application. Hence, developing scalable and controllable massive fabrication method of few-layer BP with good uniformity and crystallinity will promote the wide application of BP.\\
	\end{flushleft}
	
	\begin{flushleft}
		Furthermore, very few studies on successful fabrication of few-layer 2D materials via thermal thinning were reported: mechanically exfoliated MoS$_2$ flakes were reported to be thermally etched, which resulted in second phase formation and extensive flake damage \cite{baktiautama2013layer}$^,$ \cite{wu2013layer} . In addition, deformation and reconstruction during the sublimation of graphene have also been observed \cite{huang2009situ}. Up to now, according to our best knowledge, Huang et.al\cite{huang2014evaporative} showed successful preparation of large ($>$ 10 $\mu$m$^2$) and thin (1-2 nm) 2D Bi$_2$Se$_3$ and Sb$_2$Te$_3$ crystals by sublimation thinning. Recently, Lu\cite{lu2016thin} reported thermal thinning of tungsten telluride (WTe$_2$) flake and retain its original lattice structure. However, so far there have been no reports on any successful preparation of few-layer phosphorene or 2D materials other than Bi$_2$Se$_3$ and Sb$_2$Te$_3$ through the sublimation approach. In 2015, Liu\cite{liu2015situ} et al first observed sublimation of BP at 400 $^\circ$C starting from the formation of eye-shaped cracks along $<$001$>$ directions by in-situ transmission electron microscopy (TEM). Su\cite{su2015temperature} et al also observed the BP decomposition and sublimation at 350$^\circ$C. In 2016, another work\cite{fortin2016dynamics} indicated the sublimation of BP took place at 375$^\circ$C and the process involved detachments of pairs of P atoms. However, so far no studies have been reported using sublimation as a top-down thinning method to prepare few-layer black phosphorus.\\
	\end{flushleft}

	\begin{flushleft}
		The work was organized following the order: we first reported the layer-by-layer uniform sublimation of bulk BP flakes (starting thickness at $\sim$ 100 nm) in the temperature range from 500 K ($\sim$ 230 $^\circ$C) to 600 K ($\sim$ 330 $^\circ$C) in high purity nitrogen gas ambient; we reported the successful preparation of a 5-nm-thick crystalline BP with an area of 180 $\mu$m $^2$; we investigated thinning rates of BP at 500 K and 550 K, respectively; we recorded Raman peak intensity ratios of BP A$_g^2$ mode and underlying Si ($\frac{I_{Si}}{I_{A_g^2}}$) acquired at both high temperatures (500 K and 550 K) and room temperature as functions of BP thickness. At last, we reported bilayer phosphorene with good crystallinity was successfully prepared via sublimation thinning method. Therefore, our results introduced a facile and controllable top-down method towards uniform and crystalline few-layer BP, provided a reliable recipe evaluating BP thickness both at high temperature and room temperatures, and indicated the working temperature of future BP-based device should be lower than 500 K.  
	\end{flushleft}
	\par
	\section{Experimental methods}

	\subsection{Thermal sublimation} 
	\begin{flushleft}
		BP flakes were mechanically exfoliated from bulk BP crystals (99.998 $\%$, Smart Elements, Vienna, Austria) in a glove box using Nitto SPV 224 R tape and transferred to a pre-cleaned Si (100) wafer (4 inch, 0.56 mm thick) with a 300 nm thick thermal-grown SiO$_2$ layer; maximum temperature limit: 400 $^\circ$C). The wafer was then cut into small pieces (4 mm$ \times $4 mm). Isothermal joule heating in nitrogen gas protection (Praxair, industrial grade, purity$:$ 99.995$\%$) was performed with a Linkam TS1200 heating stage. One small piece was placed on a sapphire plate at a time in the chamber of the heating stage with nitrogen gas  flowing through the chamber. The heating stage was mounted on the motorized XY stage of a Horiba Scientific LabRAM HR800 con-focal Raman microscope. The ramp-up / cool-down rates were 100 K/min. 
	\end{flushleft}

	\subsection{Raman measurement}

	\begin{flushleft}
		All Raman measurements were performed on a Horiba LabRam HR800 Raman system in a backscattering configuration with 441.6 nm excitation. Spectra were collected through an Olympus 50X (NA=0.55) objective lens and recorded with 2400 lines mm$^{-1}$ grating which has a spectral resolution of 0.27 cm$^{-1}$. When performing Raman measurement at high temperature, the laser power was miniaturized by adding filter to be less than 0.2 mW/$\mu$m$^2$, so that the heating effect induced by the laser power can be minimized. The acquisition time and accumulation times were optimized with 6 sec and 2 times, respectively, in order to get a signal-to-noise ratio of 100 approximately as well as avoid long time radiation damage to BP samples. Moreover, in this work, all Raman maps were acquired using following configurations: the increments of and Y direction were 1.6 and 1.5 $\mu$m /step, respectively. To note that, the laser spot size was $\sim$ 2 $\mu$m$^2$. In addition, during Raman mapping measurements, all BP flakes were placed in the heating stage protected with high purity nitrogen gas purged in to avoid oxidation. 
	\end{flushleft}

	\subsection{Determination of crystal orientations of BP and underlying Si (100) substrate} 
	\begin{flushleft}
		The 441.6 nm incident laser beam was polarized, and an analyzer was placed in parallel configuration before the entrance of spectrometer in the same manner like previous study \cite{kim2015anomalous}$^,$\cite{ling2016anisotropic}$^,$\cite{luo2016study} . BP samples were placed on a rotation stage and rotated 180$^\circ$ about the microscope optical axis in 12 steps (15$^\circ$$/$step). Considering the periodicity of A$_g^1$ and A$_g^2$ modes are 180$^\circ$, and the maximum Raman intensity of A$_g^1$ and A$_g^2$ modes are orthogonal, the maximum Raman intensity of A$_g^1$ or A$_g^2$ mode could be easily measured with rotating only 90$^\circ$ (6 steps).  In this work, the sample was rotated at angle with the maximum Raman intensity of A$_g^1$ mode. What's more, since the rotation angle acquired from previous 6-step operation might be still rough from the real angle with maximum Raman intensity of A$_g^1$ mode, half-interval search method was used for further refinement to get more precise angle. Similar operations were directed for underlying Si(100) to find the maximum and minimum Si Raman intensity. In addition, during every step, laser spot was focused at the same point on sample to  ensure the consistency of results. 
		
	\end{flushleft}
	
	\subsection{Atomic Force Microscopy (AFM)}

	\begin{flushleft}
		AFM measurements were performed using contact mode by an Asylum Research Molecular Force Probe 3D atomic force microscope and a Bruker Atomic Force Microscopy System. Samples were stored in an enclosed container fulfilled with desiccants and high impurity nitrogen gas before transferring to AFM measurements in order to minimize the oxidation. 
	\end{flushleft}

	\section{Results and discussion}
	\subsection{Uniform color changes with no observable cracks were observed during the isothermal Joule heating. }
	\begin{flushleft}
		Here, we reported uniform color changes with no observable area changes were observed during the isothermal joule heating under the protection of high purity nitrogen gas, instead of any area shrinking. Fig.1 $($A$)$$-$$($D$)$ showed the optical images of a BP sample on SiO$_2$/Si substrate at four different times during 600 K heating from 0 to 51 minutes. \\
	\end{flushleft}
	
	\begin{flushleft}
		Obvious changes of color, which is directly related to the thickness, were observed. In Fig.1 $($A$)$, the pink zone $($area 2$)$ was thinner than the green zone $($area 1$)$. Fig.1 $($B$)$ showed area 1 turned to red and area 2 to peach after heating for 21 minutes. In Fig.1 $($C$)$, area 2 turned to gray at 41 minutes, which is the characteristic color of few-layer phosphorene as reported previously \cite{castellanos2014isolation}$^,$\cite{lu2014plasma} . At 51 minutes, area 2 was fully sublimated and no Raman peak could be detected. During the full process, no significant area changes were observed, and the colors of each region were uniformly changed, which suggested that the sublimation process was uniform and happened layer by layer. 
	\end{flushleft}
	
	\begin{flushleft}
		In addition, in Fig.2, high temperature annealing (500 K and 550 K) were performed at 2 pieces of BP flakes. Spatial Raman maps and thickness measurements by atomic force microscopy (AFM) of both BP flakes were taken before and after high temperature annealing. (A) and (D) were optical images of a 36-nm-thick BP flake reduced to 13-nm-thick after annealing at 500 K. In comparison with the Raman maps of A$_g^1$ and A$_g^2$ modes of the pristine 36-nm-thick flake shown in (B) and (C), those of the flake after annealing process for 120 minutes shown in (E) and (F) indicated that obviously, the BP sample kept its integrity and uniformity after heating process, which confirmed the isotropic layer-by-layer sublimation of BP.
		Similarly, the comparison between (G)-(J) and (K)-(M) of another flake also indicated that the layer-by-layer exfoliation of annealing process operated at 550 K for 25 minutes. Moreover, we noticed the color changed with thickness, which could be explained by interference effect between light and different thickness of BP flake.
	\end{flushleft}

	\begin{flushleft}
		However, previous  studies\cite{liu2015situ}$^,$\cite{su2015temperature}$^,$\cite{fortin2016dynamics} all reported BP sublimation and decomposition were above 350 $^\circ$C, so was the oxidation of BP caused by trace impurities such as water and oxygen in nitrogen gas during annealing, that lowered the sublimation temperature? here we eliminated the oxidation of BP as the main factor facilitating sublimation. Favron\cite{favron2015photooxidation} first indicated that light, oxygen and water were the three factors cause the oxidation of BP; Luo\cite{luo2016surface} reported that with 5$\%$ oxygen/Ar or 2.3$\%$ H$_2$O/Ar, the oxidation rate is $<$ 5$\AA$ for 5h from XPS results; Wang\cite{wang2016degradation} reported BP could retain stable in pure water without the presence of oxygen molecules from nuclear magnetic resonance (NMR) spectroscopy results; Li\cite{li2016role} reported that annealing could remove the metastable oxygen adsorbed on the surface of BP. Hence according those studies on oxidation mechanism of BP, as well as considering protection of high purity (99.995$\%$) nitrogen gas in this work, we attributed the self-separation of BP was the main reason for the sublimation.\\ 
	\end{flushleft}
	
	\begin{flushleft}
		Moreover, Du\cite{du2010ab} et al first reported theoretical prediction interlayer separation energy of BP was $\sim$ -0.06 eV with a layer distance of 5.5 $\AA$; Sresht\cite{sresht2015liquid} et al calculated the energy required to peel a mono-layer phosphorene from bulk BP, and proposed the wedges intercalation mechanism of BP separation in liquids; liu\cite{liu2015situ} et al reported edges of BP flakes curled and sublimation occurred at edges and defects at 400 $^\circ$C by using Transition Electron Microscopy (TEM); another report\cite{fortin2016dynamics} using in-situ low energy electron microscopy also confirmed those and revealed the sublimation occurred via detachments of two connected phosphorus atoms rather than single atoms. Thus, we attributed the BP sublimation process at temperatures lower than 350$^\circ$C to be: first the edges of BP curled up and increased the relative layer distance at edges, and interlayer separation energy of BP decreased; then because of relative larger thermal expansion along armchair direction\cite{liu2015situ}$^,$\cite{fortin2016dynamics} , wraps of surface layer and eye-shape cracks along armchair direction generated; then this process 
		took place back and forth and caused the thermal exfoliation. To address that, similar sublimation mechanisms involving point defects and edge reconstruction were also well studied in graphene\cite{huang2009situ}$^,$\cite{jia2009controlled}$^,$\cite{huang2010situ}$^,$\cite{barreiro2012graphene} . \\
	\end{flushleft}

	\subsection{Investigation of BP thinning rates at 500 K and 550 K}

	\begin{flushleft} 
		After ascertaining the layer-by-layer separation of BP, we implemented an investigation on the thinning rates of BP at 500 and 550 K. This investigation was separated into three parts: 1. determination of crystallographic orientations of BP and underlying Si (100) substrate; 2 $\&$ 3. interrupted and continuous annealing on 4 independent BP flakes at 500 K and 550 K, respectively; 4. Raman intensity ratios $\frac{I_{Si}}{I_{A_g^2}}$ as a function of thickness.  
	\end{flushleft}

	\subsubsection{Determination of crystallographic orientations (CO) by angle-resolved polarized Raman spectroscopy (ARPRS)}
	
	\begin{flushleft}
		To address that, because of anisotropic BP and Si (100) in-plane structure, the Raman intensities depends on rotation angles. Thus identifying the crystallographic orientations of BP and Si (100) becomes a perquisite for further thickness profiling to be discussed in following session. According to previous reports\cite{kim2015anomalous}$^,$\cite{luo2016study} , A$_g^1$ and A$_g^2$ modes of BP have their maximum and minimum Raman intensities at orthogonal directions. Thus, in this work, we rotated samples A and C to have maximum A$_g^1$ and minimum A$_g^2$ Raman intensities and kept their positions while performing isothermal joule heating in heating stage. 
	\end{flushleft}
	
	\begin{flushleft}
		In Fig.3 (A), a single flake (named "Sample A") has been rotated to the right angle with maximum and minimum Raman intensities of A$_g^1$ mode and A$_g^2$ mode shown in Fig.3 (B). There are 3 areas in sample A showing with various colors indicating different thicknesses of those 3 areas. The Raman spectrum in Fig.3 (B) was taken at labeled area 1. Similarly, another sample B was operated for CO determination by ARPRS in the same manner like sample A.  
	\end{flushleft}

	\begin{flushleft}
		What's more, the angle-dependent polarized Raman response of underlying Si (100) is a sinusoidal function with a period of 90$^\circ$  \cite{tuschel2012raman}$^,$\cite{yoo2015polarized} . In consideration of penetration depth and further investigation in Raman intensity ratios $\frac{Si}{A_g^2}$ as a function of thickness, we preferred the rotation angle difference between maximum Raman intensity of A$_g^1$ mode and that of Si (100) should be within 15$^\circ$, in other words, when BP sample is with maximum Raman intensity of A$_g^1$ mode, more than 50$\%$ of maximum Raman intensity of the underlying Si (100) can still be measured. In Fig.3 (C), the intensities of BP A$_g^1$ and A$_g^2$ modes as well as underlying Si (100)
		substrates were plotted as functions of rotation angles. The rotation angles of maximum A$_g^1$ and Si intensity showed a difference of 15$^\circ$.
	\end{flushleft}
	
	\subsubsection{Interrupted annealing $\&$ Continuous annealing at 500 K and 550 K}
	\begin{flushleft}
		The interrupted annealing was operated as: 5 cycles of annealing process were directed on single BP flake at 500 K (or 550 K). The annealing time  for each cycle were illustrated in Fig.4 (B). Thicknesses before and after each annealing process were acquired by AFM measurements, which were demonstrated in Fig.4 (C). The characteristic colors of different thickness were recorded as well. The continuous annealing was operated as: one cycle of annealing process was directed on single BP flake at 500 K (or 550 K), thicknesses before and after annealing process were acquired by AFM measurements. In addition, the color-thickness map was built up and shown in Fig.4 (C), Fig.4 (F), Fig.6 (C), Fig.6 (G) and Fig.(7). 
	\end{flushleft}
	
	\begin{flushleft}
		In Fig.4 (C), it illustrated that, for each annealing cycle, each area was reduced same thickness, revealing that different areas underwent isotropic sublimation rates. After that, in Fig.4 (D) - (F), sample B with 2 different areas was annealed at 550 K for 6 cycles and the annealing time for each cycle was listed in Fig.4 (E),  the recorded thickness changes were plotted in Fig.4 (F), revealing that, at the annealing temperature of 550 K, sublimation took place isotropically in different areas. 
	\end{flushleft}
	\begin{flushleft}
		Moreover, in Fig.4 (C), after annealing for 3 cycles, the thickness of area 3 (labeled in  Fig.4 (A)), was reduced to $\sim$ 5 nm (shown in Fig.5 (A)). In Fig.5, the AFM image (B), and spatial Raman maps of A$_g^1$ (D) and A$_g^2$ (E) modes indicated the uniformity and crystallinity of area 3. The area of this 5-nm-thick flake was around 180 $\mu$m$^2$. Especially, in Fig.4 (C),  Raman intensities of A$_g^1$ and A$_g^2$ remains strong while the Si Raman intensity stiffens in comparison with the Raman spectrum of original sample showed in Fig.3 (C). Since the critical thickness of BP from bulk to 2-dimensional was predicted as $\sim$ 10 nm\cite{tran2014layer} (or 20 layers), the annealing process could effectively provide the capability of thinning down thick BP flake to few-layer BP, namely phosphorene. Furthermore, PL spectra were collected at area 3 of sample A by using 633 nm and 441.6 nm lasers, which were listed in Fig.5 (F) and (G), respectively. For excitation wavelength of 441.6 nm, strong PL peak appears at 1353 nm; for 633 nm laser, PL peak shows at around 1335 nm. In general, measured results indicate the PL position of the BP sample is at around 1350 nm, which is in accordance with the PL position of 3L BP in previous report\cite{zhang2014extraordinary}. Thus, the real thickness of area 3 of sample A is around 2 nm. Surface oxidation (see bubbles on BP sample in Fig.5 (B)) during AFM measurement might be responsible for deviation.   
	\end{flushleft}
	\begin{flushleft}
		In order to probe into the accurate thinning rates, continuous annealing process were performed on two thick BP flakes at 500 K and 550 K, respectively. In Fig.6 (A) and (B), the thickness of sample C was reduced from 85 to 45 nm after annealing process at 500 K for 221 minutes. The gradual color changes were recorded (images were included in Figure 1 of supporting information). In addition, in Fig.6 (B) and (F), black spots on sample C and D were caused by frequent laser exposure (Raman measurements) during annealing process. The thickness of each color were read from the color-thickness map and then interpolated in the profile shown in Fig.6 (C). In Fig.6 (E) and (F), the thickness of sample C was reduced from 105 to 45 nm after annealing process at 550 K for 51 minutes (images of color changes were included in Figure 2 of supporting information). Similarly, the thickness-time profile of sample D annealed at 550 K was also plotted in Fig.6 (G). In Fig.6 (C) and (G), the thickness-annealing time profiles of sample A and C for interrupted annealing was also integrated. Obviously, the thickness-annealing time profiles of interrupted and continuous annealing at 500 and 550 K matched. The thinning rates were $\sim$ 0.18 nm/min at 500 K and $\sim$ 1.5 nm/min at 550 K.      
	\end{flushleft}
	
	\begin{flushleft}
		In addition, the time-dependent in-situ Raman spectra at 500 K and 550 K were also taken which were shown in Fig.6 (D) and (H). To mention that, no CO determination by ARPRS of sample C and D for continuous annealing was performed, which meant the rotation angle could be arbitrary. Thus in Fig.6 (D) and (H), BP Raman modes A$_g^1$, B$_{2g}$, A$_g^2$ and Si peak could be seen. With increasing time during annealing, the Si peak stiffened while all 3 BP peak remained unchanged.    
	\end{flushleft}	
	
	\subsubsection{Raman peak intensity ratios $\frac{I_{Si}}{I_{A_g^2}}$ as a function of BP thickness}
	\begin{flushleft}
		As mentioned above, the Si Raman intensity stiffens with decreasing BP thickness (see comparison between Fig.3 (C) and Fig.5 (C), and in-situ time-dependent Raman spectra in Fig.6 (D) and (H) and BP Raman peaks of A$_g^1$ and A$_g^2$ retain. In the other side, as the CO determination has been done for sample A and B, considering no decrease of thickness during ramp up and cooling down process, and with known thickness before and after annealing cycle, we plotted Raman intensity ratios of underlying Si over BP A$_g^2$ mode ($\frac{I_{Si}}{I_{A_g^2}}$) as a function of BP thickness. In Fig.7 (A), both room temperature profiles of sample A and B match, and the ($\frac{I_{Si}}{I_{A_g^2}}$) ascends with decreasing thickness; meanwhile, high temperature profiles of sample A (500 K) and B (550 k) were given. To address that, these two profiles first provide the capability of on-site determining BP thickness at 500 K and 550 K respectively, when the rotation angle difference of maximum Si Raman intensity and BP A$_g^1$ mode is 15 $^\circ$. Moreover,for future investigations on BP thinning using annealing method, the Raman intensity ratio ($\frac{I_{Si}}{I_{A_g^2}}$) could be used as an effective way for on-site no-contact thickness determination with known rotation angle difference of maximum Si and BP A$_g^1$ Raman intensity.  
	\end{flushleft}
	
	\par
	\subsubsection{Bilayer phosphorene }
	Moreover, another cycle of heating was then performed at sample A. In Fig.8 (A) and (B), Area 3 was extirely sublimated and circled position of area 1 was reduced to $\sim$ 2-nm-thick. In Fig.5 (C)-(E), Raman spectrum at circled position of Fig.5 (A) and spatial Raman maps of A$_g^1$ and A$_g^2$ modes indicate the crystallinity of BP sample. In addition, PL spectra (See supplementary Figure.3) collected at circled area using 441.6 nm laser show PL peak position at around 900 nm, which match the PL peak position of bilayer phosphorene in previous report\cite{zhang2014extraordinary}.  
	
	\section*{Conclusion}
	In summary, this work first reported the observation of the layer-by-layer isotropic sublimation of black phosphorus at 500 - 600 K by optical microscopy and spatial Raman imaging. AFM and Raman mapping images demonstrated successful preparation of an integral, uniform and crystalline 5-nm-thick BP flake with an area of 180 $\mu$m$^2$ via sublimation thinning method. The thinning rates were concluded as $\sim$ 0.18 nm/min and 1.5 nm/min at 500 and 550 K, respectively. The Raman intensity ratios of $\frac{I_{Si}}{I_{A_g^2}}$ as functions of BP thickness at room temperature, 500 K and 550 K were measured, which could be used on-site non-contact determination of BP thickness during sublimation thinning process. A BP sample with a starting thickness of $\sim$ 100 nm was succesfully thinned down to bilayer phosphorene while remained crystallinity. Overall, the controllable sublimation thinning method could provide a new approach to reduce BP thickness to few layers while keeping its crystallinity and integrity. The sublimation thinning method will be promising in further fabrication of high quality few-layer phosphorene in large scale.

	\bibliography{Reference}

	\section*{Author contributions}
	
	W.L., and G.(M).X conceived the experiment(s). W.L. prepared the samples. W.L. performed AFM measurements. W.L conducted the Raman measurements, W.L. analyzed the results. W.L. and G.(M).X led the writing of the paper, and all the authors participated in the discussion of results. The whole project was supervised by G.(M).X. All authors reviewed the manuscript. 
	
	\section*{Acknowledgement}
	
	W.L. acknowledges Dr.David Tuschel (Horiba Scientific, Edison, NJ, U.S.A) for the discussion on angle-resolved polarized Raman spectroscopy (ARPRS). W.L. acknowledges Chris Balicki (4D LABS, Simon Fraser University, Burnaby, BC, Canada) for his assistance in Atomic Force Microscopy (AFM). 
	
	\section*{Additional information}
	\par
	\vbox{}
	
	\textbf{competing financial interests: }All other authors declare no competing financial
	interests
	\par
	
	\vbox{}
	\begin{figure}[ht]
		\centering
		\includegraphics[width=\linewidth]{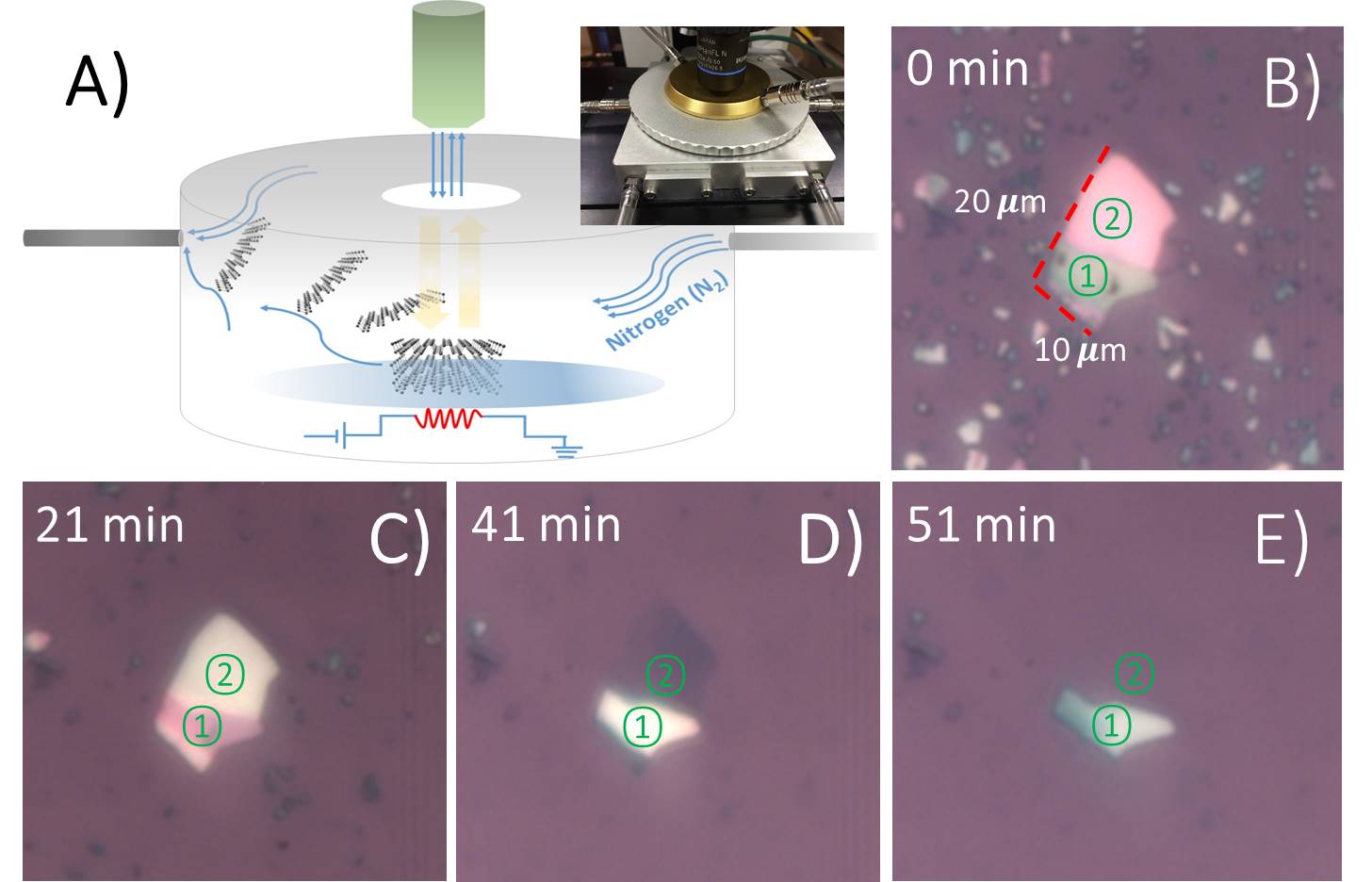}
		\caption{ (A): illustration of sublimation process of BP; (B) - (E): Observation of layer-layer-layer sublimation process of BP at 600 K. Different colors in area 1 and area 2 indicate different thicknesses. }
	\end{figure}
	\par
	\vbox{}
	
	\par
	\begin{figure}[ht]
		\centering
		\includegraphics[width=\linewidth]{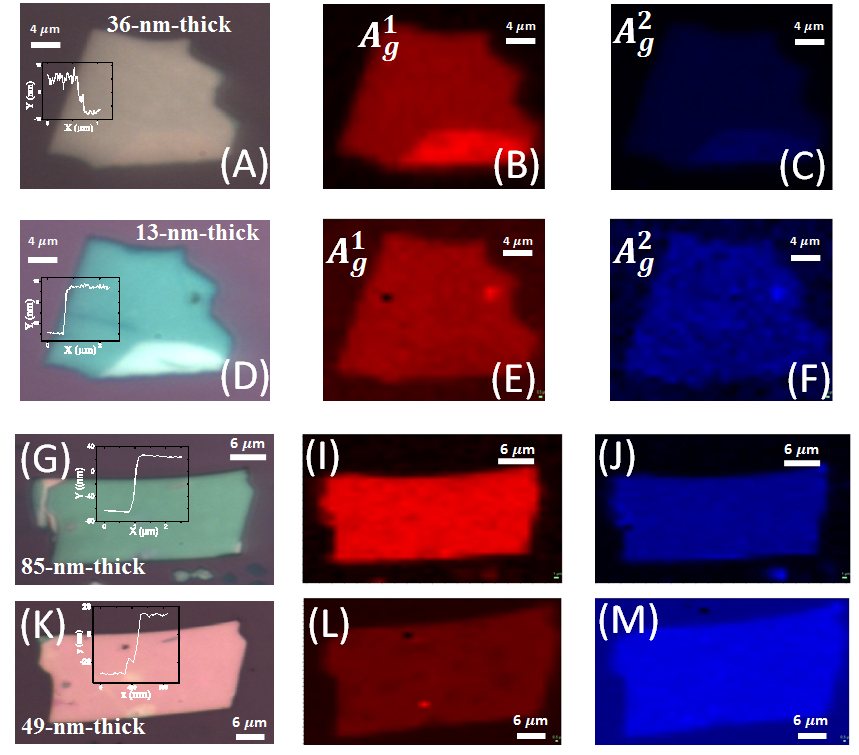}
		\caption{(A) Optical image of this BP flake before 500 K annealing process; (B) - (C): Spatial Raman maps of A$^1_g$ and A$_g^2$ peak intensity acquired on a 36-nm-thick BP flake; (D): Optical image of this BP flake after 500 K annealing process for 120 minutes, which was reduced to 13-nm-thick; (E) - (F): Raman images (A$^1_g$ and A$_g^2$ modes) of this BP flake after 500 K annealing process; (G) Optical image of this BP flake before 550 K annealing process; (I) - (J): Spatial Raman maps of A$^1_g$ and A$_g^2$ peak intensity acquired on a 85-nm-thick BP flake; (K): Optical image of this BP flake after 550 K annealing process for 25 minutes, which was reduced to 49-nm-thick; (L) - (M): Raman images (A$^1_g$ and A$_g^2$ modes) of this BP flake after 550 K annealing process;}
	\end{figure}
	
	\par
	\begin{figure}[ht]
		\centering
		\includegraphics[width=\linewidth]{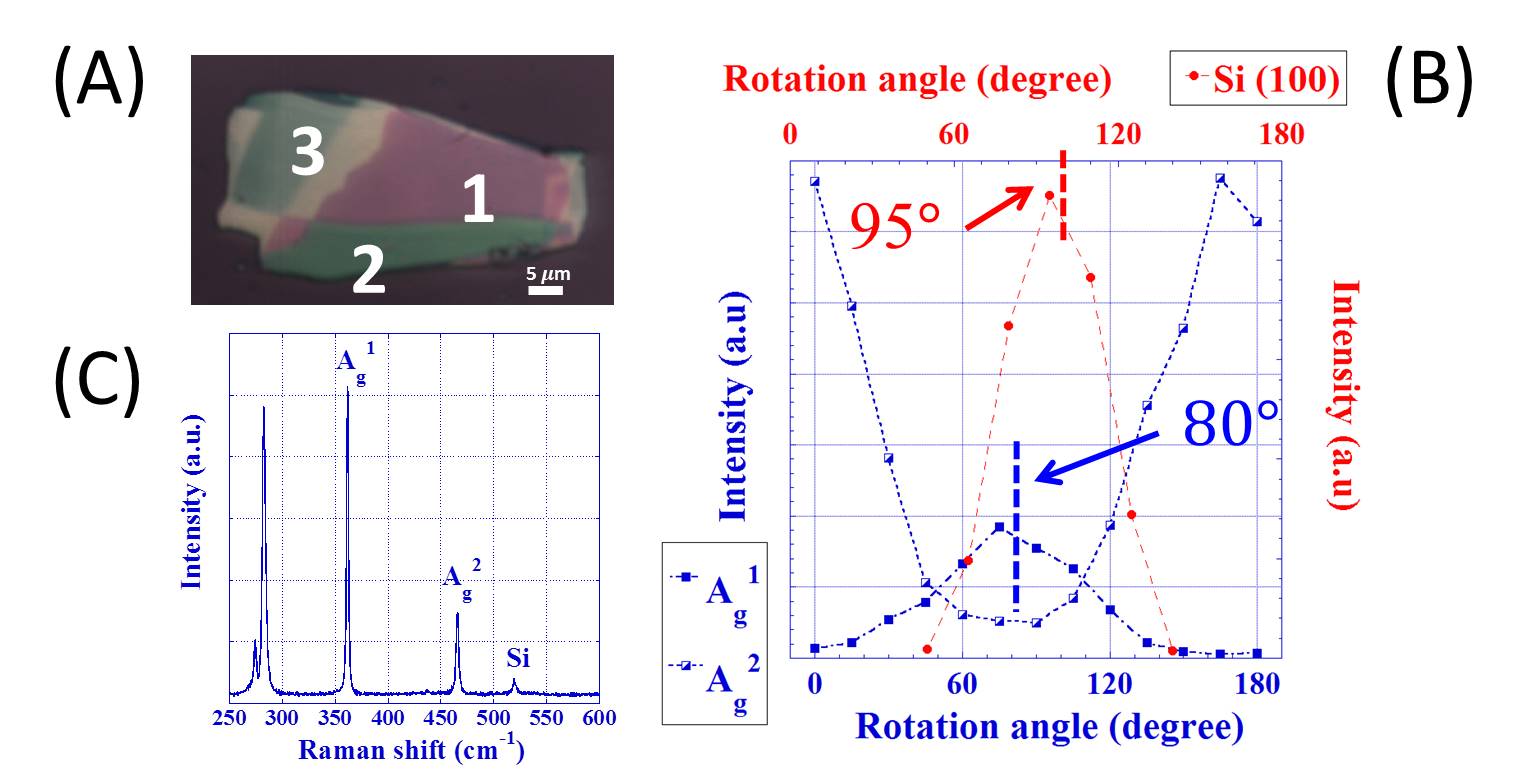}
		\caption{(A): Optical image of sample A; (B) Raman intensities of BP A$^1_g$ and A$_g^2$ modes and underlying Si(100) as functions of rotation angle. (C): Raman spectrum with maximum intensity of A$^1_g$ mode and minimum intensity of A$^2_g$ mode}
	\end{figure}
	
	\par
	\begin{figure}[ht]
		\centering
		\includegraphics[width=\linewidth]{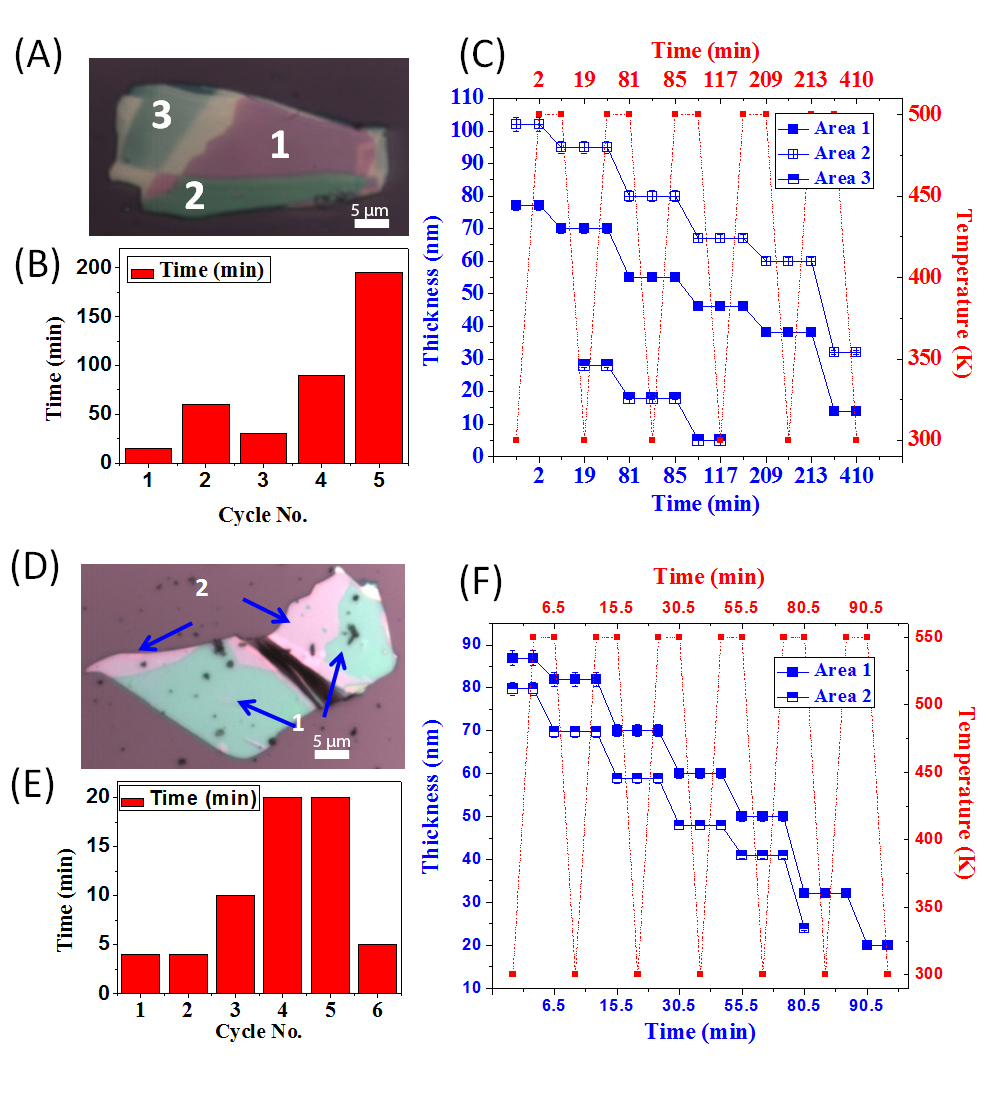}
		\caption{(A): Optical image of sample A to be heated at 500 K; (B) illustration of periodic heating process (5 cycles) to be performed at sample A; (C) Measured thickness in different areas of sample A as a function of time. (D): Optical image of sample B to be heated at 500 K; (E) illustration of periodic heating process (6 cycles) to be performed at sample B; (F) Measured thickness in different areas of sample B as a function of time.}
	\end{figure}
	
	\par
	\begin{figure}[ht]
		\centering
		\includegraphics[width=\linewidth]{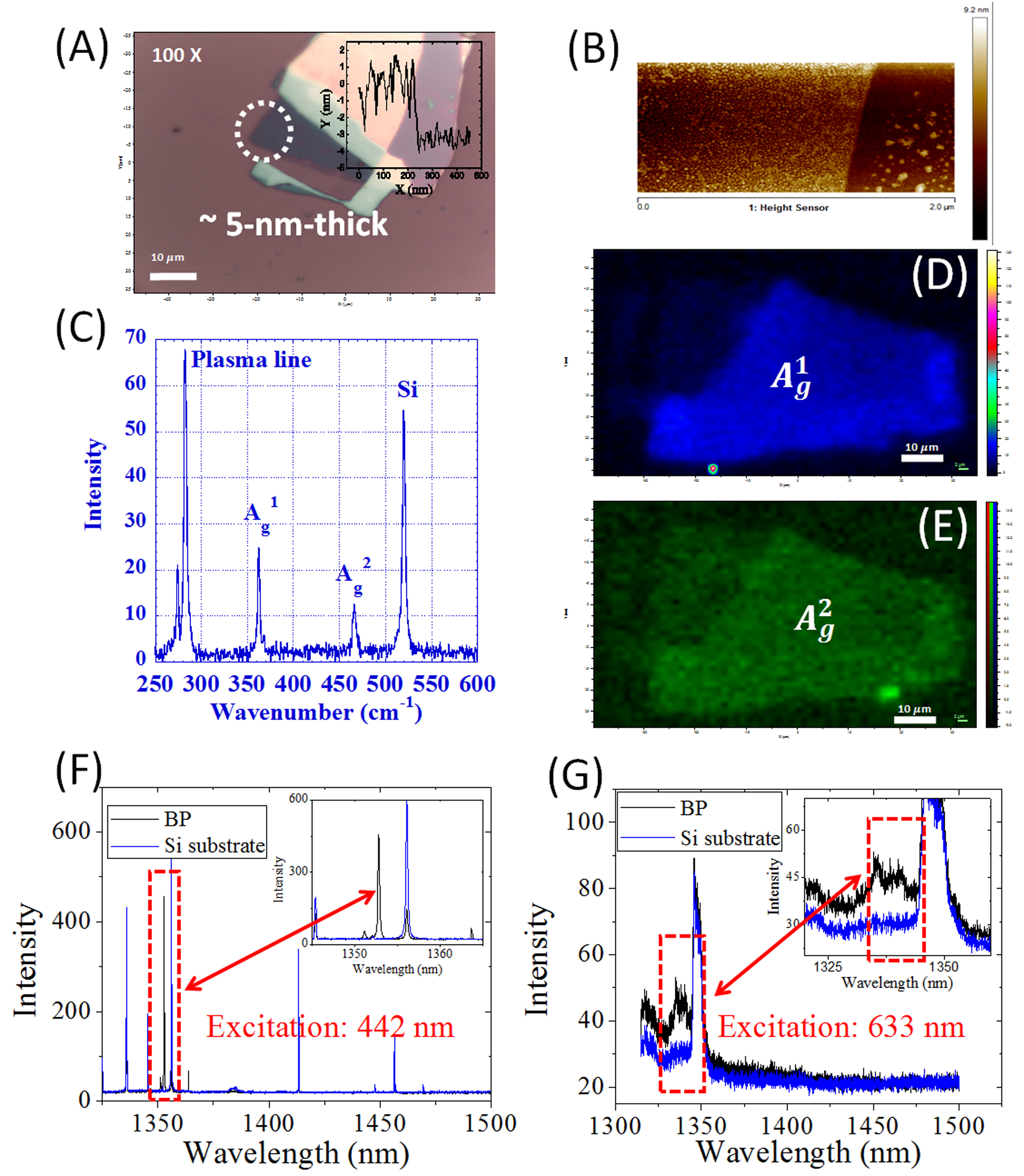}
		\caption{(A): Optical image of area 3 in Sample A, inset AFM profile indicates it has been thinned down to 5-nm-thick; (B) AFM image of area 3 of sample A; (C) Raman spectrum indicates its crystallinity; (D) $\&$ (E): Spatial Raman maps of A$_g^1$ and A$_g^2$ peak intensity modes demonstrate the integrity and crystallinity of area 3. }
	\end{figure}
	
	\par
	\begin{figure}[ht]
		\centering
		\includegraphics[width=\linewidth]{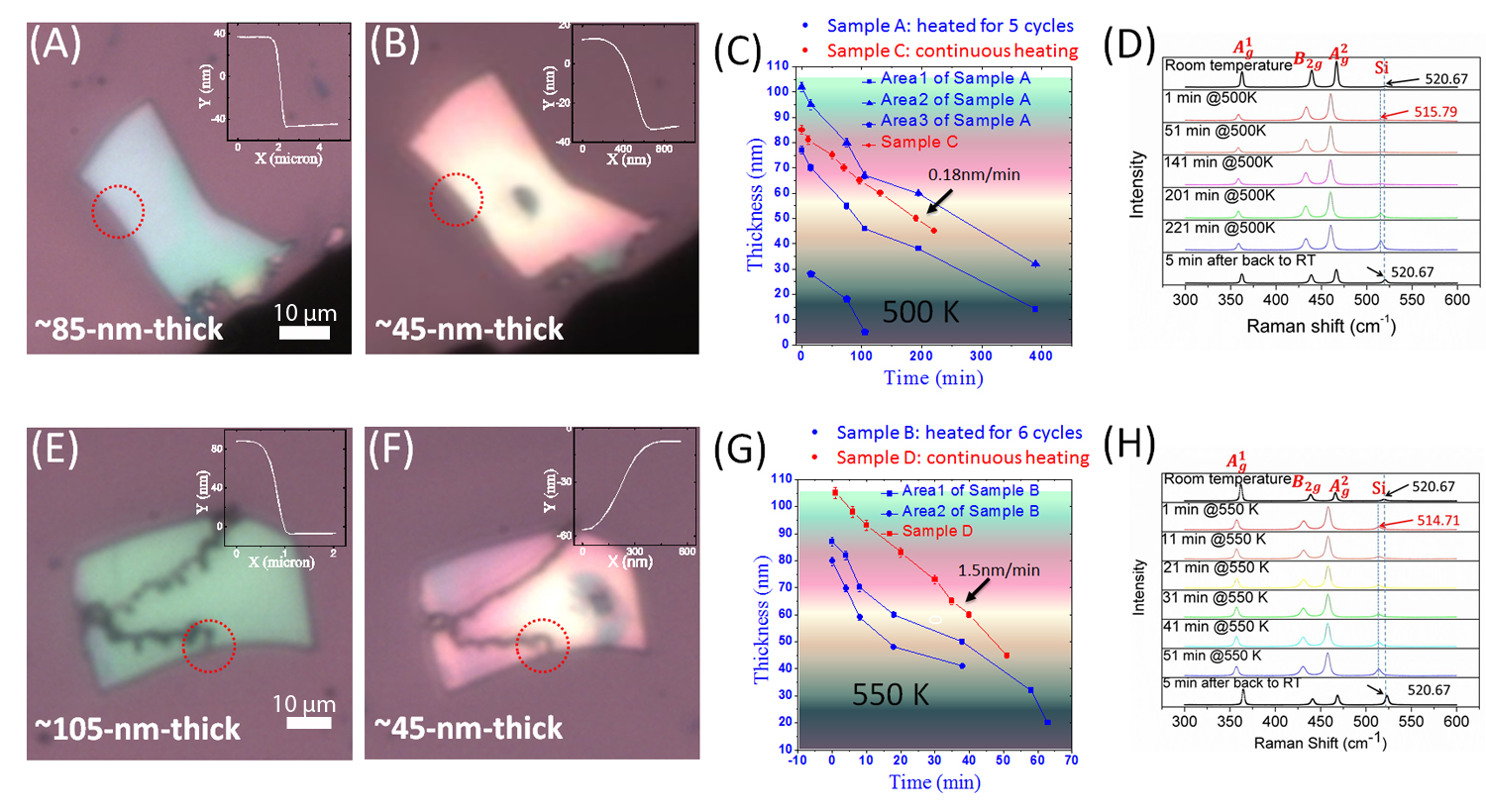}
		\caption{(A) $\&$ (B): Optical images of Sample C before and after continuous heating process at 500 K, the annealing time was 221 minutes; (C): Comparison of thinning rates of sample A and C; (D) $\&$ (E): Optical images of Sample D before and after continuous heating process at 550 K, the annealing time was 51 minutes; (F): Comparison of thinning rates of sample B and D;    }
	\end{figure}
	
	\par
	\begin{figure}[ht]
		\centering
		\includegraphics[width=\linewidth]{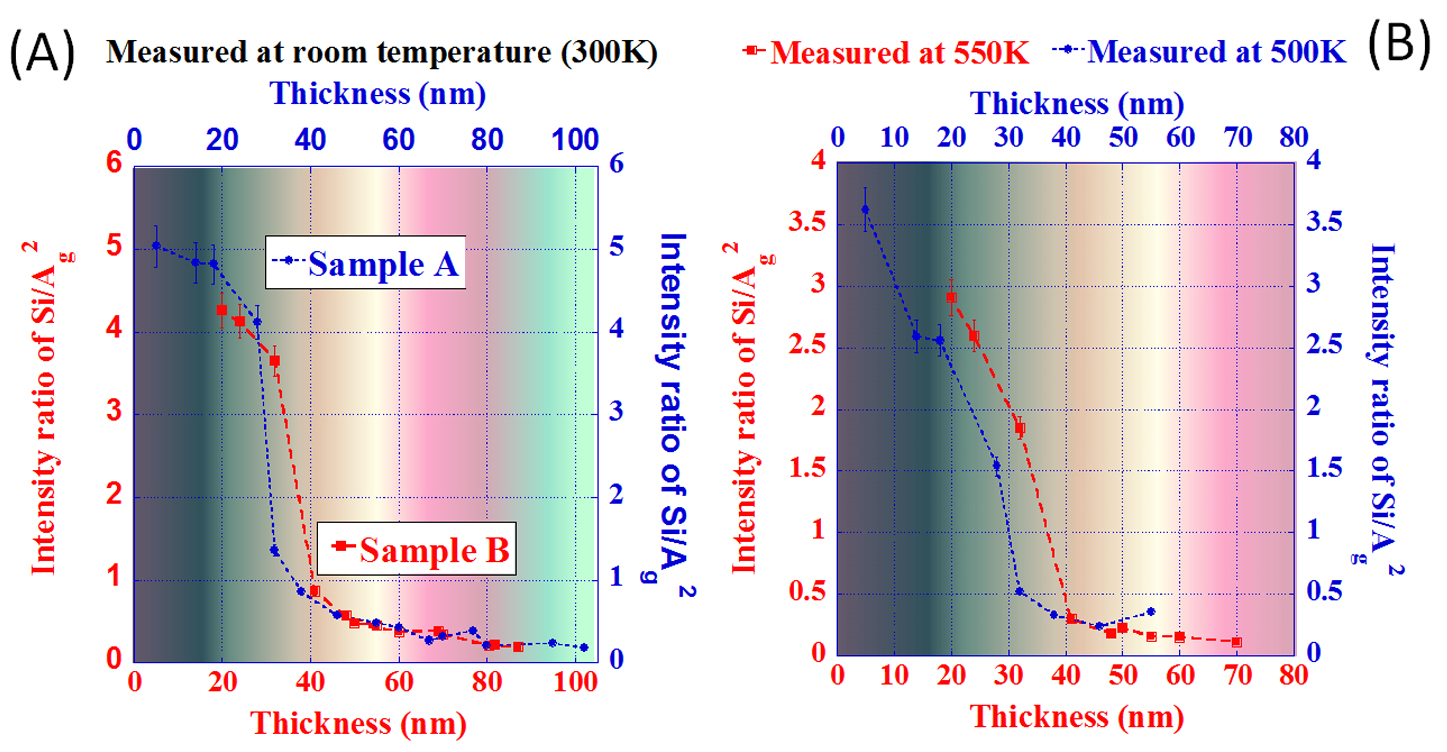}
		\caption{(A): Room temperature Raman intensity ratios of $\frac{Si}{A_g^2}$ as a function of measured thickness, here the blue line represents sample A and red line represents sample B; (B) High temperature Raman intensity ratios of $\frac{Si}{A_g^2}$ as a function of measured thickness, here the blue line represents sample A treated at 500 K and red line represents sample B at 550 K.}
	\end{figure}
	
	\par	
	\begin{figure}[ht]
		\centering
		\includegraphics[width=\linewidth]{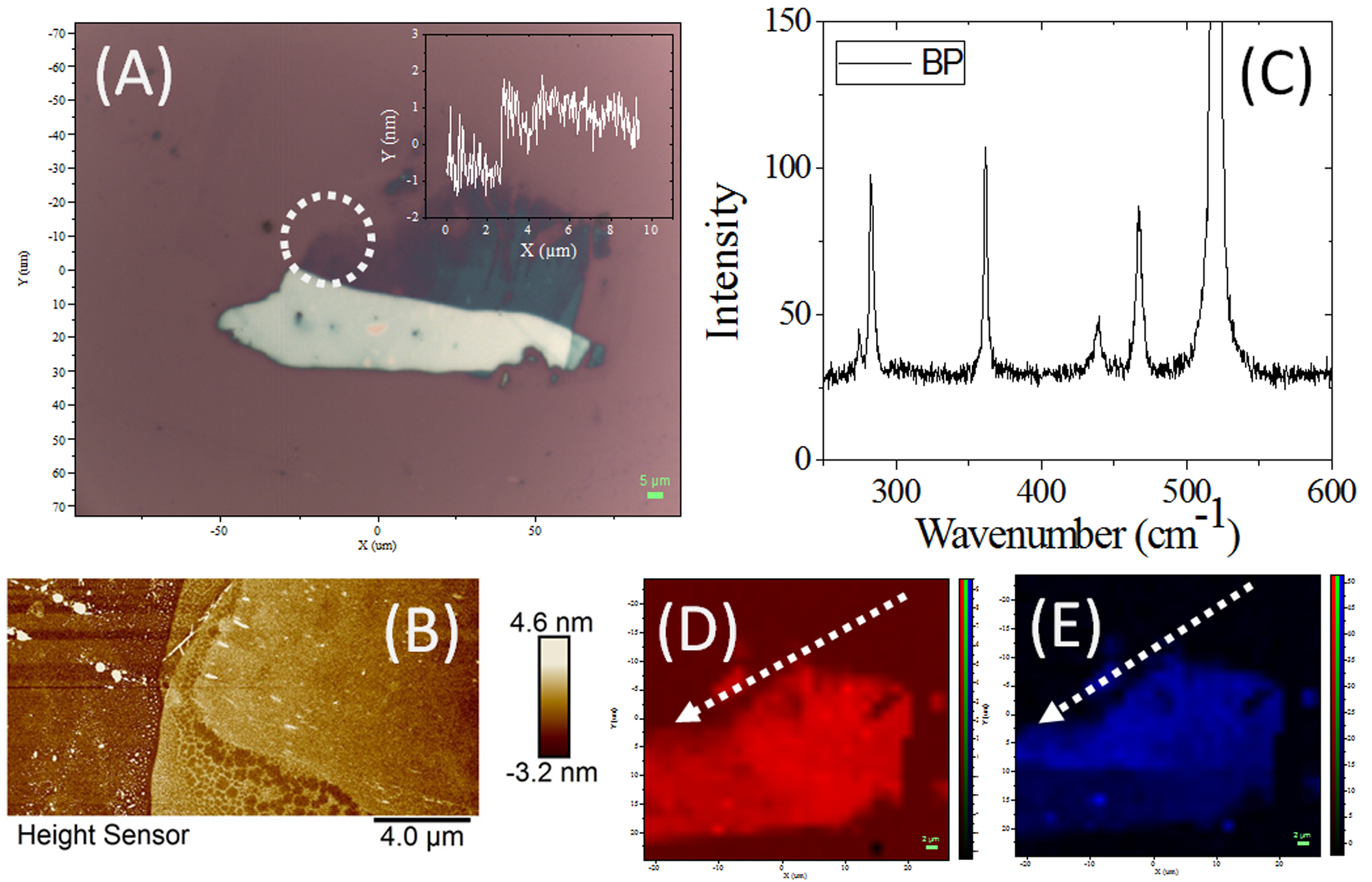}
		\caption{Sample A. (A): optical image of area 3 of sample A; (B): Raman spectrum of circled area in (A); (C): AFM image of circled area in (A); (D) \& (E): Raman maps of A$_g^1$ and A$_g^2$ modes of sample A.}
	\end{figure}

	\clearpage
	
	\centering{\large{Supplementary Information}\\}
	
	\begin{flushleft}
		1. Supplementary Figure 1:\\
	\end{flushleft}
	
	\begin{figure}[ht]
		\centering
		\includegraphics[width=\linewidth]{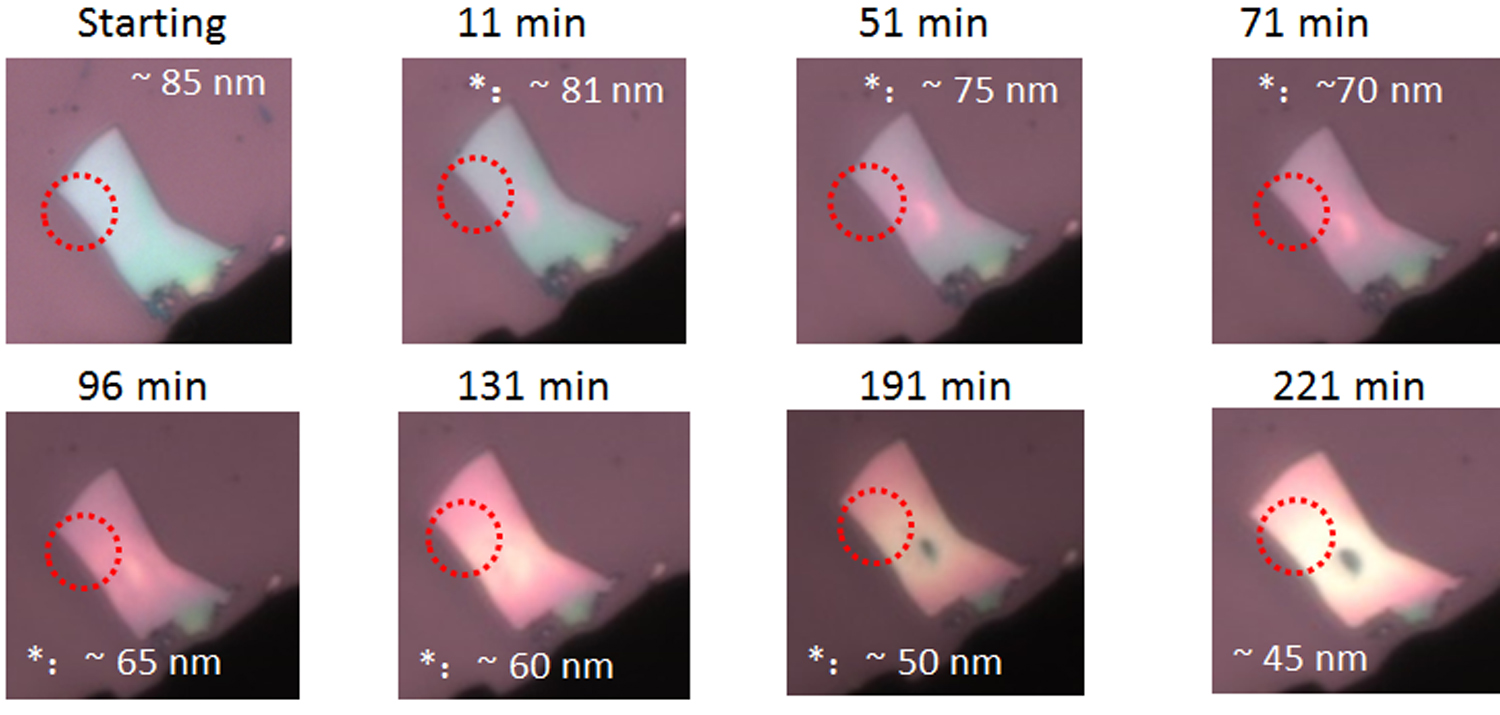}
		\caption*{Supplementary Figure 1: sample C was annealed at 500 K for 221 minutes.}
	\end{figure}
	\newpage
	\begin{flushleft}
		2. Supplementary Figure 2: \\
	\end{flushleft}
	
	\begin{figure}[ht]
		\centering
		\includegraphics[width=\linewidth]{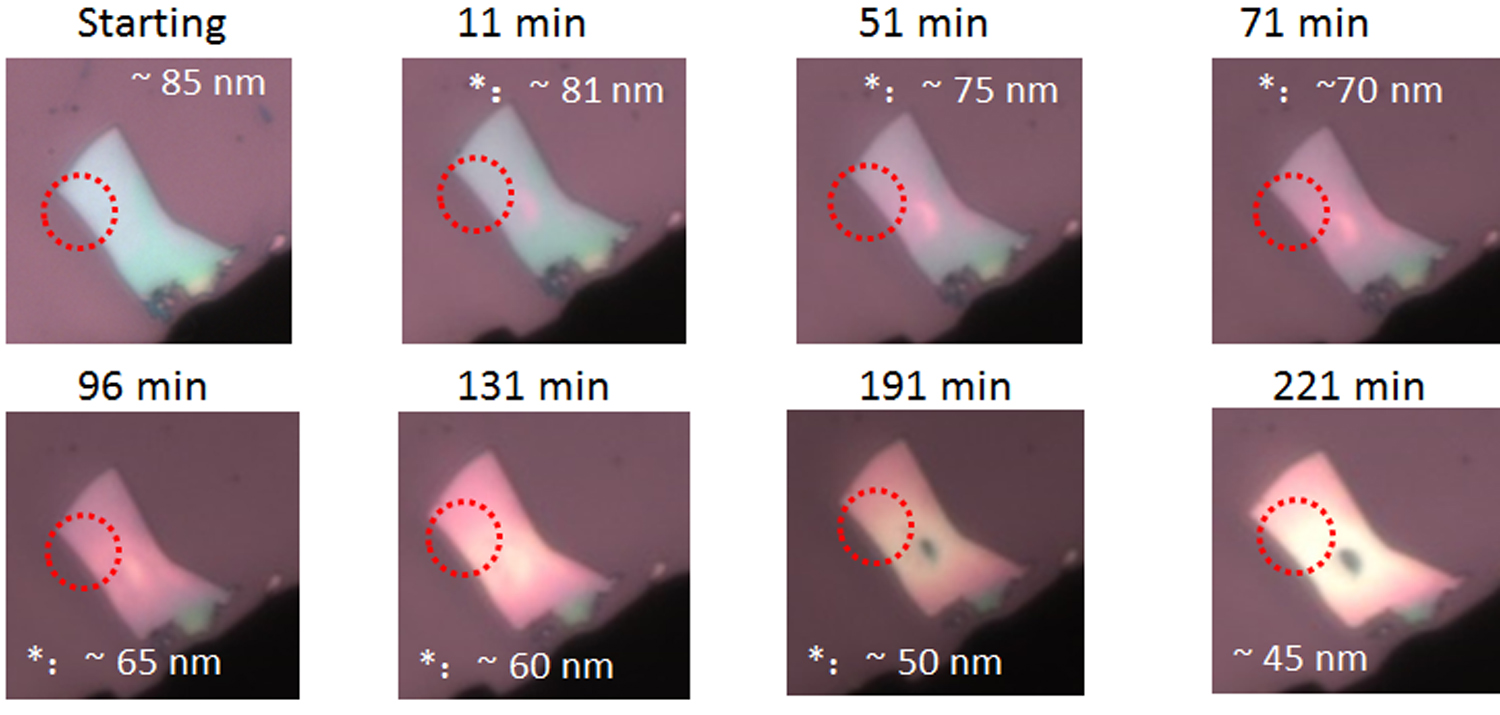}
		\caption*{Supplementary Figure 2: sample D was annealed at 550 K for 51 minutes.}
	\end{figure}
	
	\newpage
	\begin{flushleft}
		3. Supplementary Figure 3: \\
	\end{flushleft}
	
	\begin{figure}[ht]
		\centering
		\includegraphics[width=\linewidth]{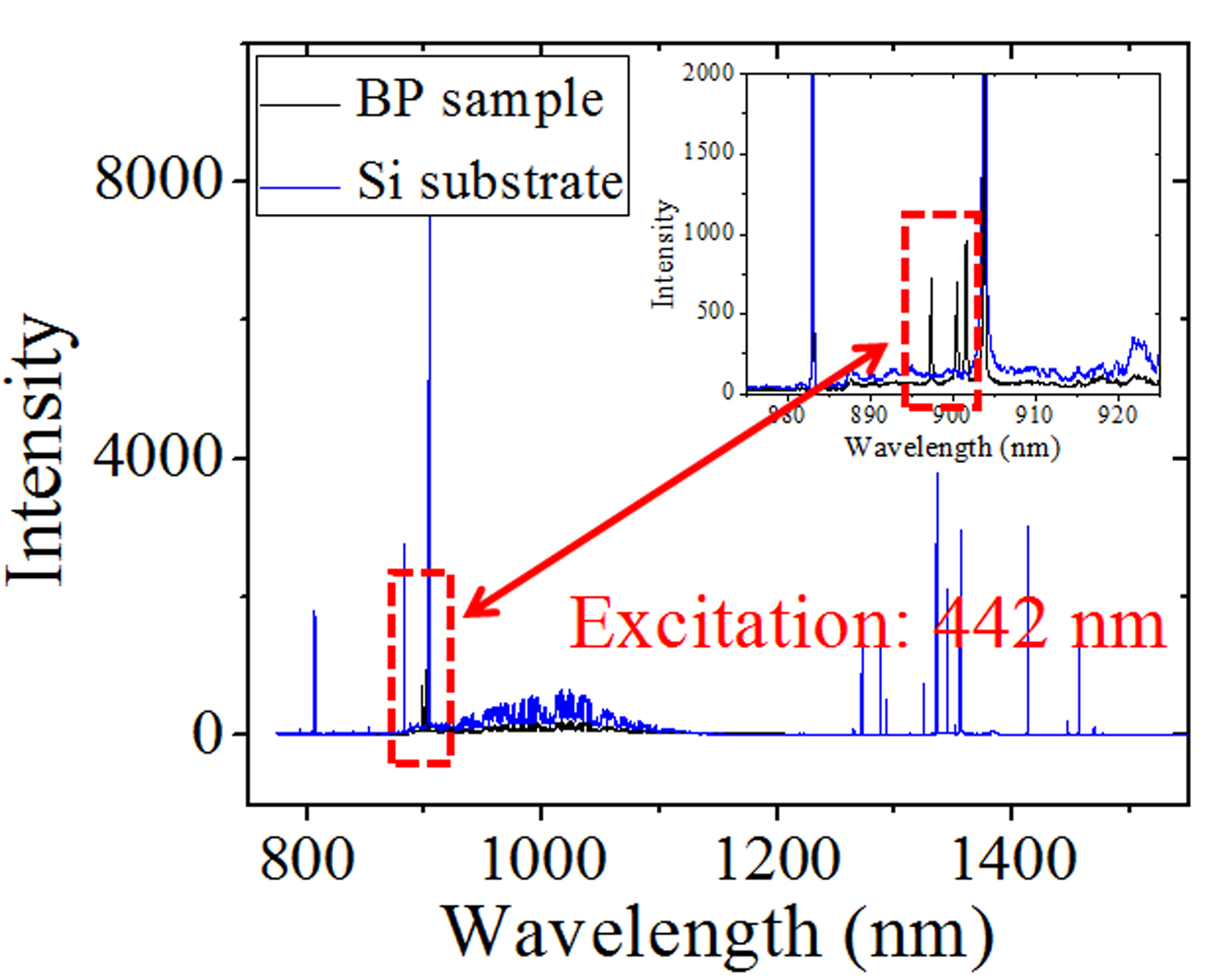}
		\caption*{Supplementary Figure 3: PL spectrum of circled area of sample A showed in Fig.8 (A). }
	\end{figure}

\end{document}